\documentclass[]{article}

\usepackage[margin=1in]{geometry} 
\usepackage{multicol} 
\usepackage{indentfirst} 
\usepackage{graphicx} 
\usepackage{amsmath}
\usepackage{float}
\usepackage{pdfpages}
\usepackage{subfigure}
\usepackage[labelfont=bf,justification=justified]{caption}
\usepackage{amssymb}
\usepackage[affil-it]{authblk}
\usepackage{latexsym}
\usepackage{amsfonts}
\usepackage{amsthm}
\usepackage{hyperref}
\usepackage{wrapfig}

\usepackage{letltxmacro}
\LetLtxMacro{\originaleqref}{\eqref}
\providecommand{\keywords}[1]{\textbf{\textit{Key words---}} #1}
\usepackage[sorting=nyt,style=numeric-comp]{biblatex}
\begin{document}

\title{A Numerical Study of Landau Damping with PETSc-PIC}
\author[1]{Daniel S. Finn}
\author[2]{Matthew G. Knepley}
\author[3]{Joseph V. Pusztay}
\author[4]{Mark F. Adams}
\affil[1,2,3]{University at Buffalo, The State University of New York}
\affil[4]{Lawrence Berkeley National Laboratory}
\date{\today}

\maketitle

\begin{abstract}
We present a study of the standard plasma physics test, Landau damping, using the Particle-In-Cell (PIC) algorithm.  The Landau damping phenomenon consists of the damping of small oscillations in plasmas without collisions.  In the PIC method, a hybrid discretization is constructed with a grid of finitely supported basis functions to represent the electric, magnetic and/or gravitational fields, and a distribution of delta functions to represent the particle field.  Approximations to the dispersion relation are found to be inadequate in accurately calculating values for the electric field frequency and damping rate when parameters of the physical system, such as the plasma frequency or thermal velocity, are varied.  We present a full derivation and numerical solution for the dispersion relation, and verify the PETSC-PIC numerical solutions to the Vlasov-Poisson for a large range of wave numbers and charge densities.
\end{abstract}


\section{Introduction}
In 1936, Lev Landau first formulated a simple kinetic model, now referred to as the Fokker-Plank equation in Landau form or simply just the Landau equation, for the description of charged particles in a plasma performing Coulomb collisions \cite{Landau1936}.  Ten years later, Landau furthered this discovery by predicting the damping of non-relativistic, collisionless plasma oscillations, or Langmuir waves, for the first time \cite{Landau1946}.  The basic concept proposed in that paper, that a conservative phenomenon exhibits irreversible behaviors, has since influenced hundreds of papers and become one of the foundational problems in plasma physics.  Thus, the phenomenon is now referred to as \textit{Landau damping}.  In his seminal paper, Landau used the solution to the Cauchy problem for the linearized Vlasov-Poisson equation around a spatially homogeneous Maxwellian equilibrium.  Landau solved the equation analytically using Fourier and Laplace transforms and concluded that the electric field damps exponentially and that the decay is a function of the wavenumber, $k$, of the perturbation.  In~\cite{BohmGross1949}, Bohm and Gross provide a simple explanation for the damping in plasmas.  In essence, plasmas exhibit a tendency to remain approximately field free.  Therefore, if electric fields are introduced, either by external disturbance or by an incomplete space charge neutralization, the newly introduced fields will be forced out by a reaction from the free charges.

Through the years, numerous others have extensively examined Landau damping in literature \cite{VanKampen1955,Jackson1960,Dawson1961}.  In 2009, a rigorous solution to the nonlinear Vlasov-Poisson equation was given by Villani and Mouhot in~\cite{Villani2009}.  In their paper, the damping phenomenon is reinterpreted in terms of transfer of regularity between kinetic and spatial variables, rather than exchanges of energy, with phase mixing being the driving mechanism.

Developed in parallel to the theory behind Landau damping, numerical methods for approximating solutions to the kinetic plasma system were pioneered by Vlasov~\cite{Vlasov1938}. The Particle-In-Cell (PIC) method has been a popular choice for numerically simulating plasmas since its inception~\cite{Harlow1955,Harlow1967}, as it can considerably reduce the complexity of the system in comparison to a direct $N$-body methods. The PIC method is a hybrid discretization algorithm comprised of two separate sets of bases for evaluation of different aspects of the problem. These bases are the particle basis, where the particle is represented by some (usually radially symmetric) shape function, and the mesh basis, where a mean field approach may be taken to computing different field quantities from external and self consistent forces.  Typically, the continuum field solve is handled by employing the finite element method, although other formulations have used splines~\cite{Cheng1976}, finite difference methods~\cite{Banks2019}, etc.

In this paper, we present a Particle-In-Cell (PIC) method for solving the Vlasov-Poisson system using the Portable Extensible Toolkit for Scientific Computing (PETSc) \cite{PETScManual,petsc-web-page}.  PETSc-PIC uses symplectic integration schemes~\cite{AbhyankarBrownConstantinescuGhoshSmith2014} for particle pushing while conducting field solves with a finite element method~\cite{LangeMitchellKnepleyGorman2015,KnepleyBrownRuppSmith13}.  The goal of PETSc is to provide composable pieces from which optimal simulations can be constructed.  PETSc user level APIs allow applications to delay implementation choices, such as solver details, until runtime using dynamic configuration~\cite{BrownKnepleySmith14}.  PETSc-PIC solvers fully conserve the moments, mass, momentum and energy, at each time step while also preserving entropy monotonicity.  Recent advances in the PETSc-PIC code~\cite{Pusztay2022} also include conservative projections between the finite element and particle basis, a key step towards hybrid FEM-particle algorithms.

\section{Problem Formulation}

Consider the Vlasov-Poisson system, a common variation of the more general Vlasov-Maxwell-Landau system of equations in the non-relativistic case where the magnetic and collisional effects are neglected.  It can be an effective model for strongly non-Maxwellian plasmas.  The Vlasov equation,
\begin{equation}
  \label{eq:Vlasov}
  \frac{\partial f}{\partial t} + \mathbf{v} \cdot \frac{\partial f}{\partial \mathbf{x}} - \frac{q_e}{m} \mathbf{E} \cdot \frac{\partial f}{\partial \mathbf{v}} = 0,
\end{equation}
describes the evolution of the phase space distribution, $f(\mathbf{x},\mathbf{v},t)$, defined over the domain $(\mathbf{x},\mathbf{v}) \in \mathbb{R}^D \times \mathbb{R}^D$ where $D$ is the spatial dimension.  The electric field is obtained using Poisson's equation,
\begin{equation}
  \label{eq:Poisson}
  \Delta \phi (\mathbf{x},t)  = -\frac{\rho}{\epsilon_0},
\end{equation}
where $\phi$ is the electric potential, $\rho$ is the charge density and $\mathbf{E}=-\nabla \phi$.  The charge density contains a neutralizing background term, $\sigma$, such that,
\begin{equation}
  \label{eq:ChargeDensity}
  \rho(\mathbf{x},\mathbf{v},t) = \sigma - q_e \int_{\mathbb{R}^D} f (\mathbf{x},\mathbf{v},t) d\mathbf{v}.
\end{equation}
This neutralizing background simulates the effect of ions on the electrons in the domain.  The use of a stationary, uniform background charge is based on the assumption that the ions are much heavier than the electrons and thus feel little influence from them.

In order to study the linear Landau damping phenomenon, we consider the initial particle distribution,
\begin{align}
  \label{eq:InitDist}
  f(x,v,t=0) = \frac{1}{\sqrt{2 \pi v_{th}^2}} e^{-v^2/2 v_{th}^2} (1 + \alpha \cos(k x)) \\
  (x,v) = \left[ 0,2 \pi / k\right] \times \left[-v_{max},v_{max}\right] \nonumber
\end{align}
where $v_{th} = \sqrt{K T_e/m}$, $\alpha = 0.01$, $k=0.5$, $v_{max} = 10$ and the boundaries are periodic.  An important piece of PIC methods for the Vlasov-Poisson system is the reduction of noise.  A primary source of noise in PIC methods can be traced to the initial discrete distribution of particles in phase space.  We mimic a ``quiet start''~\cite{Friedberg1969,Byers1970,Denavit1981} continuum initialization in this work by placing particles at the center of the spatial and velocity cells and weighting them based on the initial distribution function $f(x,v,t=0)$.  This method is also used in~\cite{Myers2016}, where particles are further remapped back to the cell centers every few steps.  The remapping step provides enough noise reduction to accurately observe nonlinear effects in damping, however, as we are, for now, concerned only with the linear case of Landau damping, we will ignore the remapping phase.

\subsection{Linear Landau Damping}
\label{sec:LandauDamping}
We seek to first derive a set of equations to understand the damping of plasma oscillations in our system and to calculate expected values for the damping rate and electric field oscillation frequency.  These expressions are found by first deriving the dispersion relation for a plasma.  The derivation shown follows from~\cite{Chen1984}.  Consider a uniform plasma with an initial distribution $f_0(v)$ with zero initial electric and magnetic fields, $\mathbf{E}_0=\mathbf{B}_0=0$.  To first order, the perturbation in $f(x,v,t)$ is denoted by $f_1(x,v,t)$ such that,
\begin{equation}
  \label{eq:Perturbation}
  f(x,v,t) = f_0(v) + f_1(x,v,t).
\end{equation}
Plugging \eqref{eq:Perturbation} in to \eqref{eq:Vlasov} gives,
\begin{equation}
  \label{eq:VlasovPert}
  \frac{\partial f_1}{\partial t} + \mathbf{v} \cdot \frac{\partial f_1}{\partial \mathbf{x}} - \frac{q_e}{m_e} \mathbf{E}_1 \cdot \frac{\partial f_0}{\partial \mathbf{v}} = 0.
\end{equation}
Assuming that the ions are massive and fixed and that the waves are one-dimensional plane waves $f_1 \propto e^{i(kx-\omega t)}$, \eqref{eq:VlasovPert} becomes,
\begin{equation}
  \label{eq:f1}
  f_1 = \frac{i q_e E_x}{m_e} \frac{\partial f_0 / \partial v_x}{\omega - k v_x}.
\end{equation}
Recall the Poisson equation \eqref{eq:Poisson}, with the potential $\phi$ replaced by the divergence of the electric field,
\begin{align}
  \label{eq:Poisson2}
  \nabla \cdot \mathbf{E} &= \nabla \cdot \mathbf{E}_1 = - \frac{\rho}{\epsilon_0} \\
  &= - \frac{1}{\epsilon_0} \left(\sigma - q_e \int \left(f_0(v) + f_1(x,v,t) \right)d v\right). \nonumber
\end{align}
With zero initial electric field, the electric field vector is replaced by the electric perturbation, $\mathbf{E}_1$, which takes the form $\mathbf{E_1} = E_x e^{i(kx - \omega t)} \mathbf{\hat{x}}$.  Furthermore, at equilibrium, the neutralizing background is equal to the total weight of the electron distribution, $\sigma = q_e \int f_0 dv$, leaving only the perturbation term $f_1$ in the Poisson equation. Thus we are left with,
\begin{equation}
  \label{eq:PoissonLD}
  i k \epsilon_0 E_x = - q_e \int f_1 dv.
\end{equation}
Substituting \eqref{eq:f1} into \eqref{eq:PoissonLD} and dividing by $i k \epsilon_0 E_x$, we have,
\begin{equation}
  \label{eq:3DDispersion}
  1 = \frac{q_e^2}{k m_e \epsilon_0} \int \frac{\partial f_0 / \partial v}{\omega - k v} dv.
\end{equation}
The integral in \eqref{eq:3DDispersion} is a three-dimensional integral, however for the Maxwellian or other factorable distribution, integration in the 2nd and 3rd dimension is simple.  Evaluating the \eqref{eq:3DDispersion} integral in the 2nd and 3rd dimension, and substituting in the plasma frequency, $\omega_p = \left( n_e q_e^2/m \epsilon_0\right)^{1/2}$, leaves the dispersion relation,
\begin{equation}
  \label{eq:DispersionRelation}
  1 = \frac{\omega_p^2}{k^2} \int^{\infty}_{-\infty} \frac{\partial f_0 / \partial v_x}{v_x - (\omega / k)} dv_x.
\end{equation}
Landau showed that this problem can be solved rigorously by means of the Laplace transform method.  Importantly, it is necessary to go around the singularity in the integrand in \eqref{eq:DispersionRelation} in the complex plane.  The solution to \eqref{eq:DispersionRelation} takes the form,
\begin{equation}
  \omega = \omega_r + i \gamma,
\end{equation}
where $\omega_r$ represents the real oscillations of the plasma and $\gamma$ the imaginary, which Landau showed to be the part of the solution driving the damping of the oscillations.  Following Landau's method~\cite{Chen1984}, an approximation for the oscillation and damping terms can be derived, given by,
\begin{align}
  \label{eq:OmegaAndGamma}
  \omega_r &= 1 + \frac{3}{2} \hat{k}^2,\\
  \gamma &= - \sqrt{\frac{\pi}{8}} \frac{1}{\hat{k}^3} \exp \left[-\frac{1}{2\hat{k}^2}\right].\nonumber
\end{align}
A normalized form of the wavenumber $k$ has been introduced to simplify the equations going forward.  The normalized wavenumber, $\hat{k}$ is given by,
\begin{equation}
   \hat{k} = \frac{k v_{th}}{\omega_p}
\end{equation}
where $v_{th} = \sqrt{K T/m}$ is the thermal velocity.  For all examples, we non-dimensionalize so that $v_{th} = 1$.  The real part of the solution to \eqref{eq:DispersionRelation} was similarly derived by Vlasov in~\cite{Vlasov1938}, however Vlasov did not account for the imaginary damping term.

These approximations are valid for the case where $\hat{k} \ll 1$ but their accuracy degrades considerably as $\hat{k}$ approaches $1$ and higher.  Even when $\hat{k}=0.5$, the calculated values for $\omega_r$ and $\gamma$ differ from the numerical results by at least $5\%$.  In~\cite{McKinstrie1999}, McKinstrie draws similar conclusions, electing to derive more accurate forms of \eqref{eq:OmegaAndGamma} by expanding $\omega_r$ in powers of $\hat{k}$,
\begin{align}
  \label{eq:McKinstrie}
  \omega_r &= 1+\frac{3}{2} \hat{k}^2+\frac{15}{8} \hat{k}^4+\frac{147}{16} \hat{k}^6, \\
  \gamma &= -\sqrt{\frac{\pi}{8}}\left(\frac{1}{\hat{k}^3}-6 \hat{k}\right) \exp \left[-\frac{1}{2 \hat{k}^2}-\frac{3}{2}-3 \hat{k}^2-12 \hat{k}^4\right].
\end{align}
These new expressions are more accurate for $\hat{k}$ up to $0.4$ but still diverge from the correct values as $\hat{k}$ increases further.  Shalaby et. al. provided further refinements to these equations in~\cite{Shalaby2017}, using a numerical fitting formula, taking the form,
\begin{align}
  \label{eq:Shalaby}
  \omega= & 1+\frac{3}{2} \hat{k}^2+\frac{15}{8} \hat{k}^4+\frac{147}{16} \hat{k}^6+736.437 \hat{k}^8-14729.3 \hat{k}^{10} \\
  & +105429 \hat{k}^{12}-370151 \hat{k}^{14}+645538 \hat{k}^{16}-448190 \hat{k}^{18}, \nonumber \\
  \gamma= & -\sqrt{\frac{\pi}{8}}\left(\frac{1}{\hat{k}^3}-6 \hat{k} - 40.7173 \hat{k}^3+3900.23 \hat{k}^5-2462.25 \hat{k}^7-274.99 \hat{k}^9\right) \nonumber \\
  & \exp \left[ - \frac{1}{2 \hat{k}^2}-\frac{3}{2}-3 \hat{k}^2-12 \hat{k}^4 - 575.516 \hat{k}^6+3790.16 \hat{k}^8 \right. \nonumber \\
    &\left. - 8827.54 \hat{k}^{10} + 7266.87 \hat{k}^{12} \right].\nonumber
\end{align}
These equations give good estimates for $\omega_r$ and $\gamma$ in the case where $\hat{k}=0.5$, which is of particular interest in this paper. In fact, the values obtained from~\eqref{eq:Shalaby} in the case where $\hat{k}=0.5$ and all other parameters ($\omega_p$,$v_{th}$,$q_e$,etc.) are assumed to be $1.0$ match those commonly listed as ``analytic solutions'' \cite{Chen1984,ZhouGuoShu2001,Myers2016}.  That being said, the accuracy of the numerical fit still decreases considerably for $\hat{k}>0.6$.

An alternate, and as we will show, more accurate way to calculate $\omega_r$ and $\gamma$ for given values of $\hat{k}$ is to find them by computing the zeros of \eqref{eq:DispersionRelation}.  This was done by Canosa in~\cite{Canosa1973} for values of $k$ ranging from $0.25$ to $2.0$ in increments of $0.05$ (see Table~\ref{dataCanosa} for a selection of values).  A comparison of the approximations by Landau, McKinstrie and Shalaby to the zero finding results from Canosa is shown in Section~\ref{sec:Results}.
\begin{figure}[h]
  \centering
  \begin{tabular}[pos]{c | c c}
    $\hat{k}$ & $\omega_r$ & $\gamma$ \\
    \hline
    0.25 & 1.1056 & -0.0021693 \\
    0.5  & 1.4156 & -0.15336   \\
    0.75 & 1.7371 & -0.46192   \\
    1.0  & 2.0459 & -0.85134   \\
    1.5  & 2.6323 & -1.7757    \\
    2.0  & 3.1891 & -2.8272    \\
  \end{tabular}
  \caption{Values for $\omega_r$ and $\gamma$ for given values of $\hat{k}$ from~\cite{Canosa1973}.}
  \label{dataCanosa}
\end{figure}

\section{PETSc-PIC}

PETSc, the Portable, Extensible Toolkit for Scientific Computation, is a well-known library for numerical methods.  It provides parallel data management, structured and unstructured meshes, linear and nonlinear algebraic solvers and preconditioners, optimization algorithms, time integrators and many more functions.  The PETSc-PIC algorithm relies on two modules to handle the particle and mesh solves simultaneously.  The first, DMPlex~\cite{KnepleyKarpeev09,LangeMitchellKnepleyGorman2015,KnepleyLangeGorman2017}, is a PETSc module for generic unstructured mesh creation, manipulation, and I/O~\cite{Hapla2021}.  It decouples user applications from the implementation details of common mesh and discretization tasks.  The other important module for this work, DMSwarm~\cite{MayKnepley2017}, provides a fully parallel solution for pure particle methods (e.g. DEM, SPH, EFG) and for particle-mesh methods (e.g. PIC, FLIP, MPM, GIMP).

We start with discussion of the particle methods in the PETSc-PIC algorithm.  A method must first be chosen to represent the particle space, and for interpolation between the mesh and particle representations. There are numerous choices in shape functions for this purpose, however in our case a simple delta function representation of particles is chosen.  Thus the approximation of the distribution function is defined in the particle space as,
\begin{equation}
  f_p = \sum_p \overrightarrow{\omega_p} \delta \left( \mathbf{x} - \mathbf{x_p} \right),
\end{equation}
where $\overrightarrow{\omega_p}$ is the vector of weights, $\mathbf{x}$ are the configuration space variables and $\mathbf{x_p}$ represents the particle position and velocity, respectively.  The finite element representation, using a function space $\mathcal{V}$, is given by the weighted sum of basis functions,
\begin{equation}
  f_{FE} = \sum_i f_i \psi_i (\mathbf{x}),
\end{equation}
where $\psi_i \in \mathcal{V}$ denotes the basis functions and $f_i$ the associated finite element coefficient.

The Vlasov equation is a linear hyperbolic equation which may be written in a simpler form,
\begin{equation}
  \label{advection}
  \frac{\partial f}{\partial t}+\mathbf{z} \cdot \nabla_{\mathbf{q}} f = 0,
\end{equation}
where $\mathbf{q} = (\mathbf{x}, \mathbf{v})$ is the phase space variable and $\mathbf{z} = (\mathbf{v}, -q_e \mathbf{E}/m)$ is the combined force.  The force term $-q_e \mathbf{E}/m$ is independent of velocity, and therefore \eqref{advection} may be written in the conservative form,
\begin{equation}
  \label{cons advection}
  \frac{\partial f}{\partial t}+  \nabla_{\mathbf{q}} \cdot (\mathbf{z}f)=0.
\end{equation}
Given this new advective form of the Vlasov equation, we can rewrite the equation for the characteristics $\mathbf{Q} = (\mathbf{X}, \mathbf{V})$,
\begin{equation}
  \frac{d \mathbf{Q}}{d t} = \mathbf{z},
\end{equation}
which reexpressed with the original phase-space variables gives,
\begin{align}
  \frac{d \mathbf{X}}{d t} &= \mathbf{V}, \\
  \frac{d \mathbf{V}}{d t} &= -\frac{q_e}{m} \mathbf{E}. \nonumber
\end{align}
Since particles follow characteristics, the Vlasov equation in the particle basis becomes
\begin{align}
  \frac{d \mathbf{x}_p}{d t} &= \mathbf{v}_p, \\
  \frac{d \mathbf{v}_p}{d t} &= -\frac{q_e}{m} \mathbf{E}. \nonumber
\end{align}

The equations of motion are stepped forward in time using structure-preserving symplectic integrators which have been well studied~\cite{Hairer2006_2}.  The electric field is solved concurrently at each step using a finite element solver, discussed in the next section.

\subsection{PETSc-FEM}

At each step in the simulation, the Poisson equation is solved using the finite element method.  The gradient of the potential, i.e. the electric field, is then interpolated across each cell at the particle locations.  The interpolated electric field is then applied to the particles in the form of the Coulomb force.

The PETSc-FEM method is abstractly formalized by the \textit{Ciarlet triple} \cite{Ciarlet1976,Kirby04}, such that a finite element is a triple $(\mathcal{T},\mathcal{V},\mathcal{V}^{\prime})$, where,
\begin{itemize}
  \item the domain $\mathcal{T}$ is a bounded, closed subset of $\mathbb{R}^d$ (for $d=1,2,3,...$) with nonempty interior and piecewise smooth boundary;
  \item the space $\mathcal{V}=\mathcal{V}(\Omega)$ is a finite-dimensional function space on $\mathcal{T}$ of dimension $n$;
  \item the set of degrees of freedom (nodes) $\mathcal{V}^{\prime} = \lbrace l_1, l_2,...,l_n\rbrace$ is a basis for the dual space, that is, the space of bounded linear functionals on $\mathcal{V}$.
\end{itemize}
The cell $\mathcal{T}$ together with the local function space $\mathcal{V}$ and the set of rules for describing the functions in $\mathcal{V}$ is the \textit{finite element}.  The discretization in PETSc is handled by the PETScFE object, which contains a PetscSpace ($\mathcal{V}$), PetscDualSpace ($\mathcal{V}^{\prime}$), and DMPlex ($\mathcal{T}$).  PETScFE supports simplicial elements, tensor cells, and some special cells such as pyramids.

In general, the finite element solve for the Poisson equation can be accomplished using the standard $H^1$ function space.  In the $H^1$ space, the weak form of the Poisson equation is,
\begin{equation}
  \int_{\Omega} \nabla \psi_i \cdot \nabla \phi = \int_{\Omega} \psi_i,
\end{equation}
where $\psi \in V$ and $V$ is the set of basis functions on the cell.  The elements are then constructed such that the basis functions are continuous across the cell boundaries.

\subsection{Conservative Projections}

To preserve the conservation laws in a PIC simulation, a method must be constructed to conservatively project between the particle and grid representations.  Weak equality of the representations,
\begin{equation}
  \int_{\Omega} \psi_i f_{FE} = \int_{\Omega} \psi_i f_P
\end{equation}
is enforced on the representations to achieve this~\cite{MaddisonFarrell2012,Pusztay2022}.  Restricting this equivalence to the finite-dimensional analogues gives the matrix-vector form,
\begin{equation}
  M f_{FE} = M_{p} f_{p},
\end{equation}
where M is the finite element mass matrix,
\begin{equation}
  M = \int_{\Omega} \psi_i \psi_j,
\end{equation}
$M_p$ is the particle mass matrix,
\begin{equation}
  M_{p} = \int_{\omega} \psi_i \delta(\mathbf{x}-\mathbf{x_p}),
\end{equation}
$f_{FE}$ is a vector containing the finite element coefficients and $f_p$ is the vector of particle weights.  The entries of $M_p$ contain evaluations of the finite element basis functions at particle locations with rows being determined by the basis function index, and columns being determined by the particle indices. Moving from the particle basis to the mesh, we must invert the finite element mass matrix, which is easily accomplished with CG/Jacobi~\cite{Wathen1987}. In the other direction, we must invert a rectangular particle mass matrix, usually with LSQR~\cite{Pusztay2022}.

\section{Numerical Results}
\label{sec:Results}

In this section, the results of this numerical study are presented.  We consider the one-dimensional (1X-1V) case of the Vlasov-Poisson system.  According to \eqref{eq:McKinstrie}, derived in Section~\ref{sec:LandauDamping}, and the zero finding data from Canosa~\cite{Canosa1973}, the damping rate should be $\gamma = -0.153$ and the frequency of oscillations should be $\omega_r = 1.416$.  All runs were conducted on a single 2.4 $\left[GHz\right]$ 8-Core Intel Core i9 processor with 64  $\left[GB\right]$ of memory.  The example code and packages/options required to run it are provided in Appendix~\ref{AppendixA}.

\begin{figure}[h]
  \begin{center}
    \includegraphics[width=1.0\textwidth]{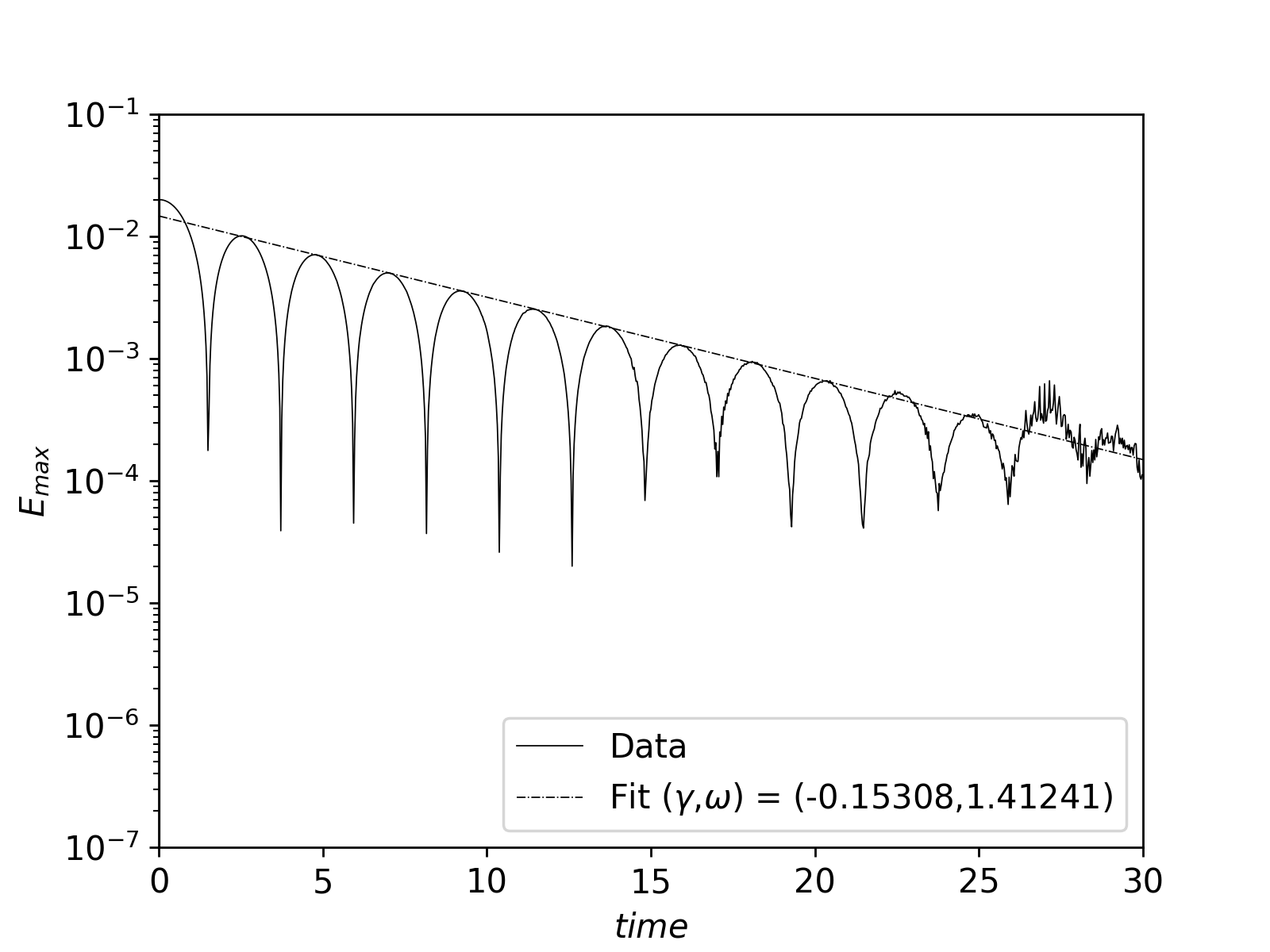}
  \end{center}
  \caption{The maximum value of the electric field as a function of time for the one-dimensional linear Landau damping problem.}
  \label{fig:LD}
\end{figure}

To begin, we show results from the densest run of the PETSc-PIC simulation with $160$ spatial cells and $8,000$ particles per cell and a PIC timestep of $dt = 0.3$. Figure \ref{fig:LD} shows the maximum value of the electric field, $E_{max} = \max_\Omega |E|$, over time.  The values for $\gamma$ and $\omega_r$ were measured by fitting the peaks of the given data.  The frequency of oscillations describes the frequency of the electric field completing one full oscillation.  Since each oscillation of $E_{max}$ is the equivalent of one half of the electric field period, we count two $E_{max}$ oscillations for each plasma oscillation.  Values achieved by the PETSc-PIC algorithm, $\gamma=-0.1531$ and $\omega_r = 1.4124$ agree within $1\%$ of the analytic values from Canosa and Shalaby et. al., which are assumed to be the most accurate for the case $k=0.5$.

\begin{figure}[h]
  \begin{center}
    \includegraphics[width=0.75\textwidth]{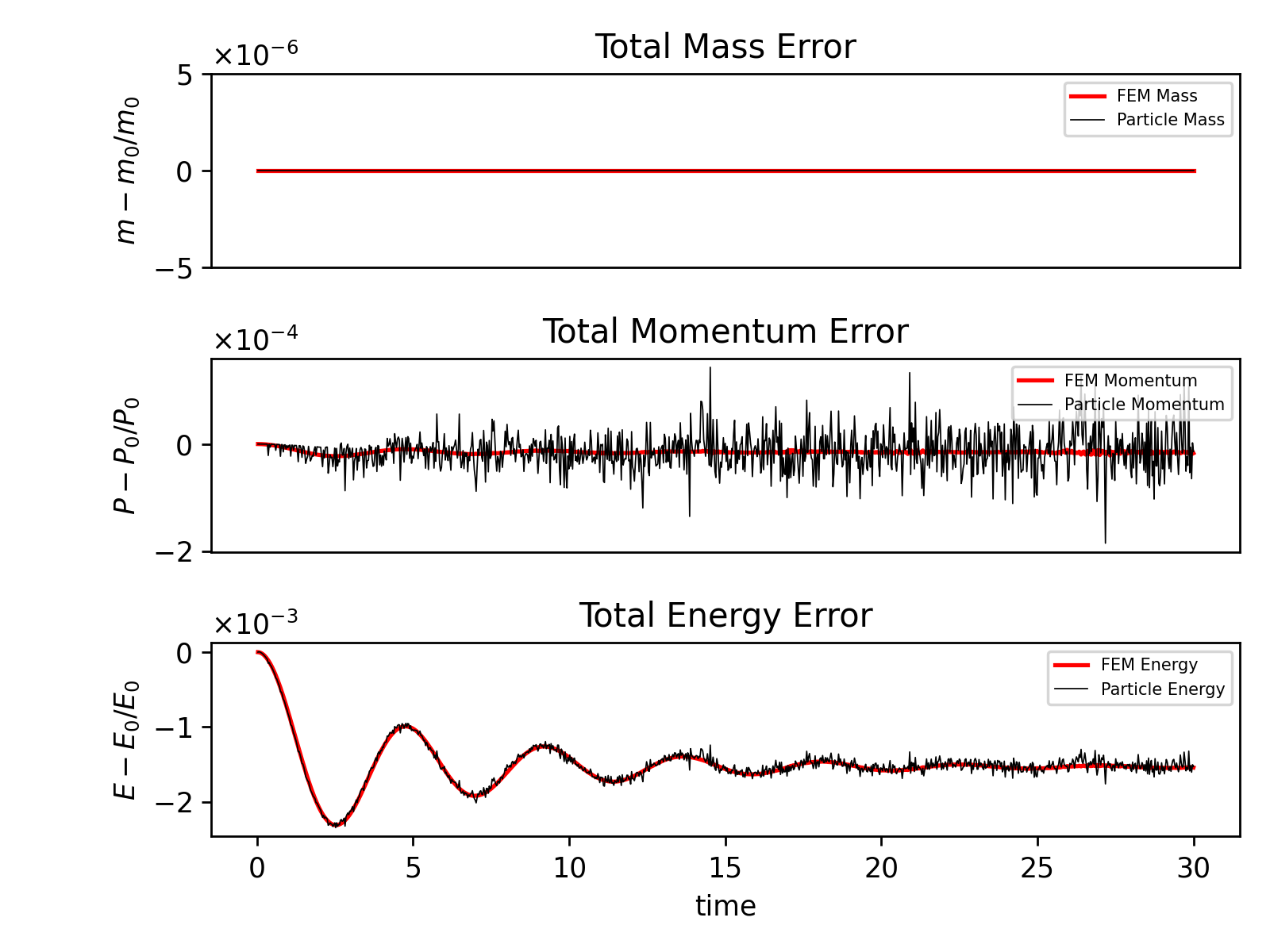}
  \end{center}
  \caption{The total mass, momentum and energy error for the particle and finite element solve.  The moment errors all converge to zero given a long enough time.}
  \label{fig:Moments}
\end{figure}

The total error in the moments, shown in Figure \ref{fig:Moments}, was shown to be stable over the entire runtime.  At early times, the error in momentum and energy fluctuate but each converges by $t=20$. This convergence comes from the use of basic symplectic integrator in PETSc-PIC which guarantees the error does not grow over time.  We also note that the error in the particle solve and the finite element solve is exactly equal, apart from an increased level of noise in the particle solve.  This confirms the effectiveness of the conservative projector used in PETSc-PIC.  More detailed tests of the conservative projector can be found in \cite{Pusztay2022}.

\begin{figure}[h]
  \begin{center}
    \includegraphics[width=0.9\textwidth]{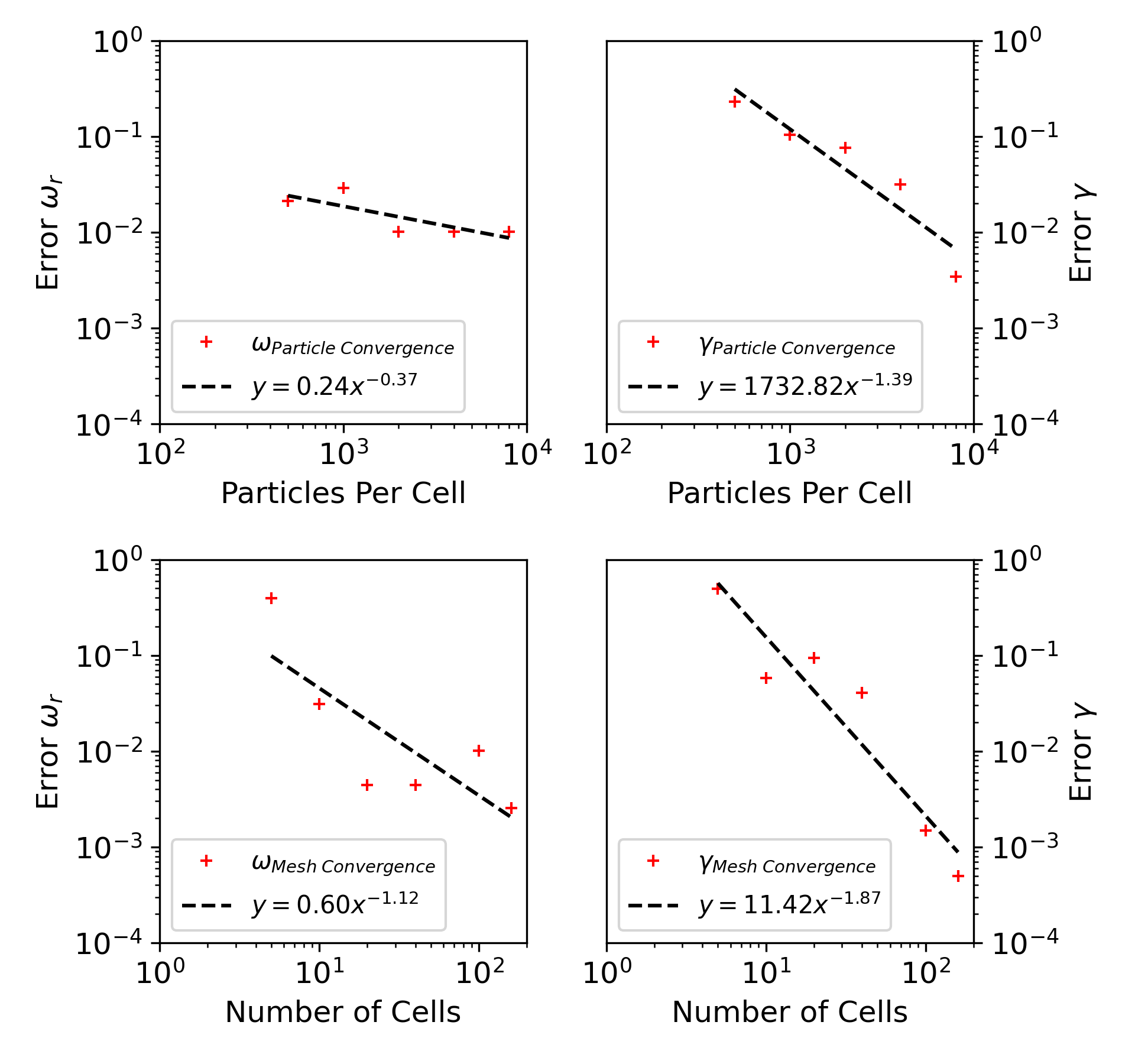}
  \end{center}
  \caption{(Top) Particle per cell number convergence plots for $N_x = 100$ and (bottom) mesh convergence plots for $N_v = 8,000$.}
  \label{fig:Convergence}
\end{figure}

Convergence studies were conducted in which either the mesh or the number of particles per cell were increased while the other was held constant.  In the case of mesh convergence, the number of particles per cell was held at $N_v = 8,000$ while in the particle number convergence tests, the number of mesh cells was $N_x = 100$.  We naively expect Monte Carlo convergence in particle number, $\mathcal{O}\left(1/\sqrt{N}\right)$, and we indeed achieve this for $\omega_r$ in the upper left of Figure~\ref{fig:Convergence}. Since we have such a regular initial particle distribution, we might hope to see Quasi-Monte Carlo convergence, $\mathcal{O}\left(1/N\right)$, and we do see this superconvergence in $\gamma$ in the upper right of Figure~\ref{fig:Convergence}. We expect $\mathcal{O}\left(h^2\right)$ convergence in the mesh resolution $h$ since this controls the error in the electric field, and we see this in $gamma$ in the lower right of Figure~\ref{fig:Convergence}. However, this convergence should quickly saturate as particle error begins to dominate, which we see in $\omega_r$ in the lower left of Figure~\ref{fig:Convergence}.

\subsection{Variations in Wavenumber and Charge Density}
\begin{figure}[h]
  \begin{center}
    \includegraphics[width=1.0\textwidth]{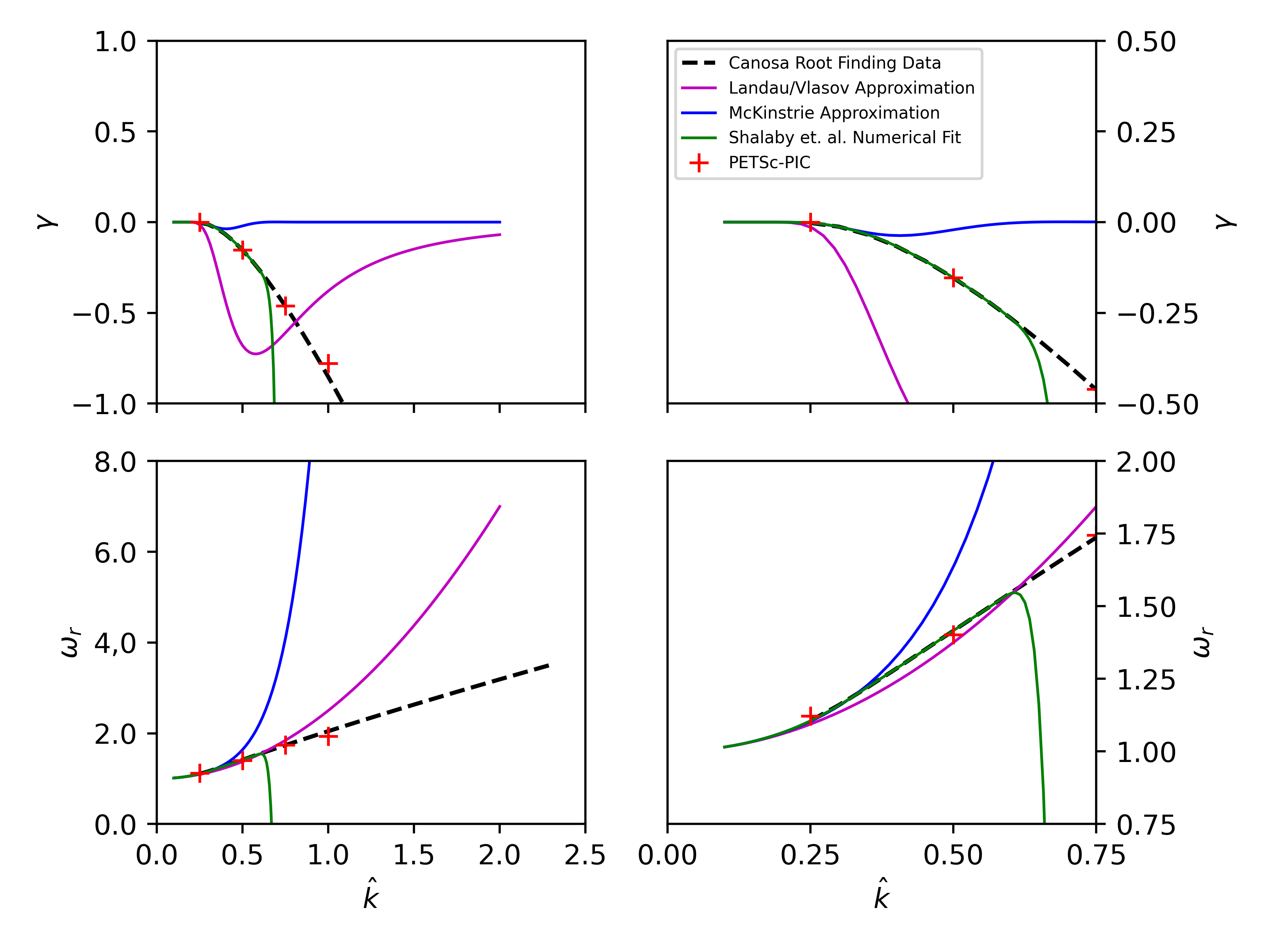}
  \end{center}
  \caption{A comparison of various approximations for $\omega_r$ and $\gamma$ to root finding results and numerical results from PETSc-PIC.  Plots on the right are zoomed in on the region $0.0 \leq \hat{k} \leq 0.75$ to show the accuracy of each approximation before they diverge from the data.}
  \label{fig:kPlot}
\end{figure}

We have thus far shown that the PETSc-PIC algorithm is an accurate and structure-preserving method for modeling plasma systems.  We next present results from tests in which the wavenumber $k$, and consequently the domain size, and the charge density were varied.  Varying either of these values impacts the value of the non-dimensional wavenumber $\hat{k}$.  In the case of the wavenumber $k$, the calculated values for $\omega_r$ and $\gamma$ were compared to the values obtained with the approximation equations~\eqref{eq:OmegaAndGamma} and~\eqref{eq:McKinstrie}, the numerical fit~\eqref{eq:Shalaby} and the zero finding results from Table~\ref{dataCanosa}.  The results from PETSc-PIC, shown in Figure~\ref{fig:kPlot}, clearly show a strong deviation of the approximation equations and the numerical fit for $\hat{k} > 0.5$ while closely matching the zero finding data.  This demonstrates that these approximations quickly break down outside of the small parameter range typically chosen in numerical studies of Landau damping.  When considering real plasma systems in which values for $k$, $\omega_p$, etc. are more dynamic, it is far more effective to use zero finding methods to calculate expected values for $\omega_r$ and $\gamma$.

\begin{figure}[h]
  \begin{center}
    \includegraphics[width=0.8\textwidth]{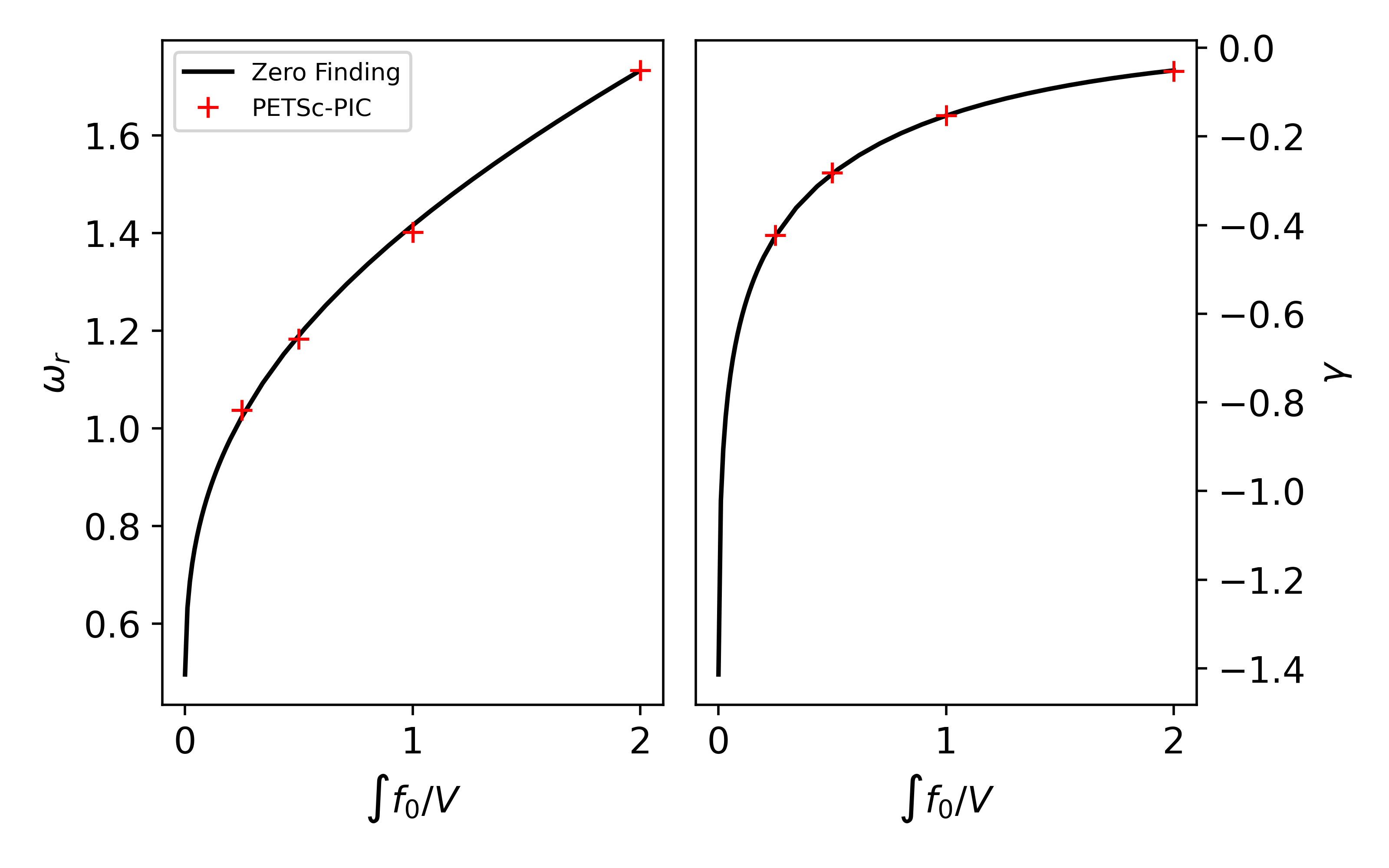}
  \end{center}
  \caption{Numerical results for varying charge densities compared to zero finding data.  The charge density is represented on the x-axis as the integral of the initial distribution over the domain volume.}
  \label{fig:CDPlot2}
\end{figure}

It may be naively assumed that data from numerical tests with varying charge densities will match the approximation equations \eqref{eq:OmegaAndGamma}, \eqref{eq:McKinstrie} and \eqref{eq:Shalaby} or even the zero finding data from Canosa, however, these analytic results are based on an assumption of unchanging charge density.  More specifically, these results are based on charge densities such that the plasma frequency, $\omega_p$, is always unity.  Therefore, to accurately compare analytic results to our data we must resolve the dispersion relation for varying charge densities.  A zero finding algorithm, using Newton's method \cite{SonnendruckerBook}, was employed to calculate new analytic values for $\omega_r$ and $\gamma$ with charge densities ranging from $0.1$ to $2.0$.  The zero finding algorithm calculates multiple values for $\omega_r$ and $\gamma$ however we select the solution containing the largest $\gamma$ which corresponds to the smallest $\omega_r$.  Other solutions found by the algorithm represent less dominant modes which can be ignored for the purposes of this study.  Figure~\ref{fig:CDPlot2} contains the results from the new zero finding algorithm along with data from numerical tests which agree perfectly.  We observe that when the charge density is increased, the frequency of oscillations also increases.  This matches the expected physical behavior of an electrically charged plasma.

\begin{figure}[h!]
  \begin{center}
    \includegraphics[width=0.7\textwidth]{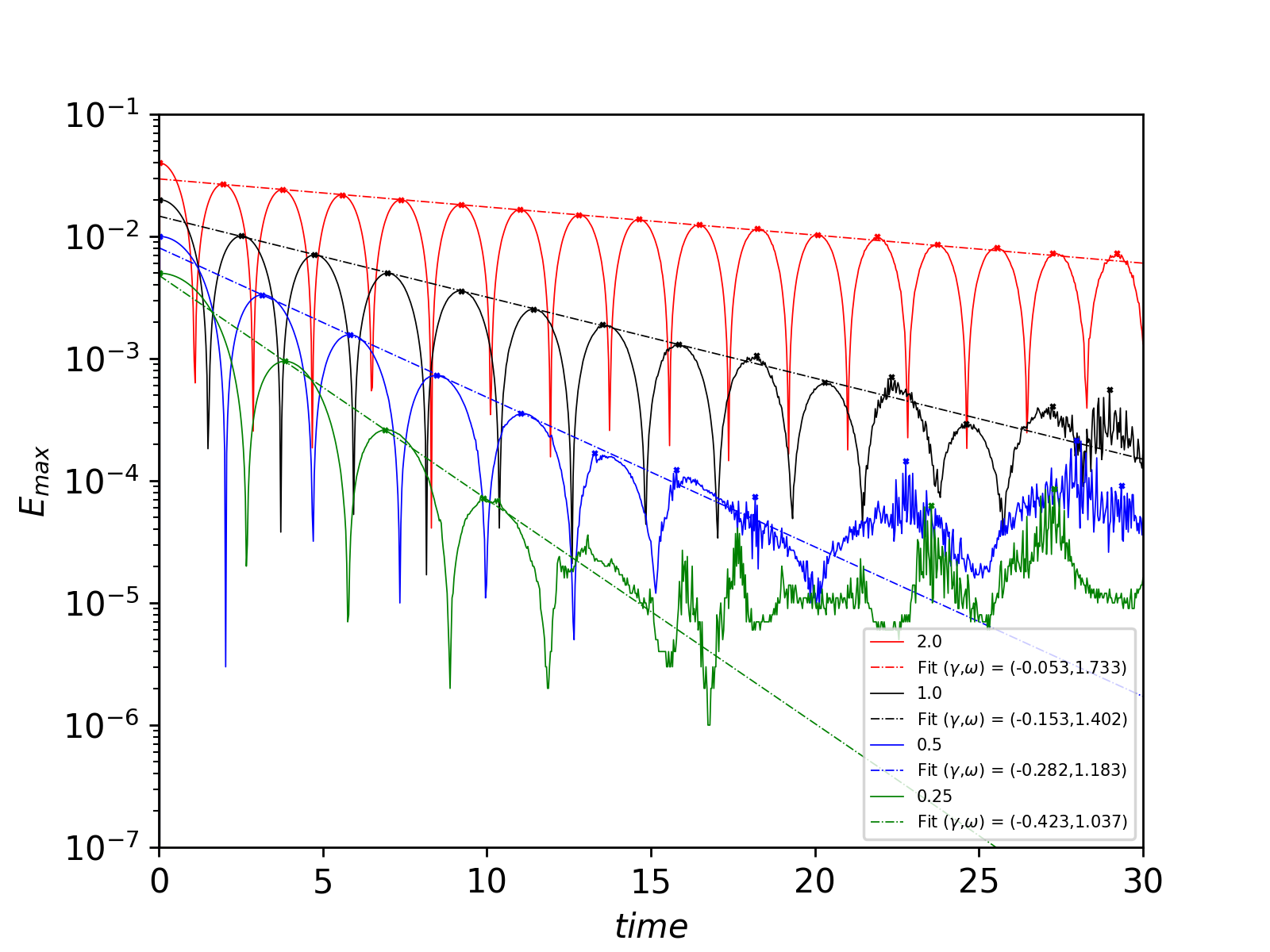}
  \end{center}
  \caption{A comparison of electric field oscillations given different charge densities.}
  \label{fig:CDPlot3}
\end{figure}

We have extended our zero finding algorithm to the case where the charge density approaches zero ($\hat{k} \rightarrow \infty$) to make note of an interesting phenomenon.  At a charge density of zero, the dispersion relation has no solution.  We capture this in Figure \ref{fig:CDPlot2}, where we observe that both $\omega_r$ and $\gamma$ are asymptotic at $\int f_0/V = 0$.  This can similarly be observed in the numerical results from our PETSc-PIC algorithm.  As the charge density is decreased, the rate at which the electric field oscillations becomes too large to resolve numerically.  In the case of $\int f_0/V = 0.25$, shown in Figure \ref{fig:CDPlot3}, we can only observe two full oscillations of the electric field before the simulation becomes too noisy.  Theoretically, the charge density could be decreased asymptotically in our simulations to observe the damping rate and frequency trends but in practice there is too much noise to resolve any real processes in the plasma.

\section{Conclusions and Future Work}

We have presented PETSc-PIC, a structure-preserving Particle-in-Cell algorithm for solving the electrostatic Vlasov-Poisson systems. The accuracy of our algorithm has been demonstrated by comparing the frequency of electric field oscillations and the damping rate of the oscillations to analytic values. We have also shown that the approximations for the frequency and damping rate break down outside of narrow ranges for the wavenumber and charge density.  These approximations are often cited in numerical Landau damping studies without further context or reference to the equations used to compute the parameters, which can lead to complications in reproducing results.  We have sought to provide a complete picture of Landau damping and the numerical methods we have used to simulate this phenomenon.

Future work with the PETSc-PIC algorithm will fall into two primary categories: improvements to the algorithm and the extension of the Landau damping test to multi-dimensional and nonlinear cases.  Improvements to the algorithm will focus on reformulation using a mixed form of the Poisson equation and $H(div)$ finite elements.  We expect that $C^0$ electric fields will decrease the noise in our particle representation over time.  While we have not observed any major negative impacts from using $H^1$ finite elements in the test problem chosen for this paper, Landau damping, using a mixed form makes a notable difference in the case of Two-Stream Instability.  PETSc currently includes support for the $H(div)$ conforming finite elements Brezzi-Douglas-Marini (BDM) and Raviart-Thomas (RT) on simplicial grids, however RT elements are currently the only element type supported on tensor cells.  We will also replicate our tests in parallel, allowing us to increase the number of particles per cell by several orders of magnitude, reducing the largest source of error in the code.

Nonlinear Landau damping is a more complex in that non-damping phenomenon, such as plasma echo, are present.  Vitally, the linearization of the Vlasov equation, used as the fundamental approximation in the study of linear Landau damping, does not guarantee that the asymptotic behavior of the linear Vlasov equation is an approximation of the asymptotic behavior of the nonlinear Vlasov equation~\cite{Villani2009}.  There are reasons to doubt that the study of the linearized equations gives any hint on the long-time behavior of the nonlinear equations.  Therefore if an algorithm is desired that can accurately capture the long-time behavior of a plasma, the nonlinear case of Landau damping must also be considered.

\appendix
\section{Appendix A}
\label{AppendixA}

The data presented in this paper can be recreated with PETSc using the DMSwarm example \textit{ex9} ($\$PETSC\_DIR/src/dm/impls/swarm/tests/ex9.c$).  Exact runtimes may vary depending on the architecture and compiler.
The DMSwarm example can be run using the following options:
\begin{verbatim}
  ./ex9 -dm_plex_dim 2 -dm_plex_simplex 0 -dm_plex_box_bd periodic,none
    -dm_plex_box_faces 10,1 -dm_view
    -dm_plex_box_lower 0,-0.5 -dm_plex_box_upper 12.5664,0.5
  -dm_swarm_num_species 1 -dm_swarm_num_particles 50
  -vdm_plex_dim 1 -vdm_plex_simplex 0 -vdm_plex_box_faces 7500
    -vdm_plex_box_lower -10 -vdm_plex_box_upper 10
  -petscspace_degree 1 -em_type primal
    -em_pc_type svd -em_snes_atol 1.e-12
  -ts_type basicsymplectic -ts_basicsymplectic_type 1
    -ts_max_time 500 -ts_max_steps 1000 -ts_dt 0.03
  -fake_1D -cosine_coefficients 0.01,0.5 -charges -1.0,1.0
    -perturbed_weights -periodic
\end{verbatim}
This example uses a 100 square cell mesh on the domain $ \left(x,v\right) \in \left[0,4\pi\right] \times \left[-10,10\right]$, with 8000 particles per cell.  A first-order basic symplectic integrator is chosen as the time integration method and $H^1$ finite elements are chosen for the field solves.

\printbibliography

@techreport{AbhyankarBrownConstantinescuGhoshSmith2014,
  title       = {{PETSc/TS}: A Modern Scalable {DAE/ODE} Solver Library},
  author      = {Shrirang Abhyankar and Jed Brown and Emil Constantinescu and Debojyoti Ghosh and Barry F. Smith},
  type        = {Preprint},
  number      = {ANL/MCS-P5061-0114},
  institution = {ANL},
  month       = {January},
  year        = {2014},
  petsc_uses  = {TS}
}

@article{Banks2019,
   author = {Banks, Jeffrey W. and Odu, Andre Gianesini and Berger, Richard and Chapman, Thomas and Arrighi, William and Brunner, Stephan},
   title = {High-Order Accurate Conservative Finite Difference Methods for Vlasov Equations in 2D+2V},
   journal = {SIAM Journal on Scientific Computing},
   volume = {41},
   number = {5},
   pages = {B953-B982},
   year = {2019},
   doi = {10.1137/19M1238551},
   URL = {https://doi.org/10.1137/19M1238551},
   eprint = {https://doi.org/10.1137/19M1238551}
}

@article{BohmGross1949,
  title = {Theory of Plasma Oscillations. A. Origin of Medium-Like Behavior},
  author = {Bohm, D. and Gross, E. P.},
  journal = {Phys. Rev.},
  volume = {75},
  issue = {12},
  pages = {1851--1864},
  numpages = {0},
  year = {1949},
  publisher = {American Physical Society},
  doi = {10.1103/PhysRev.75.1851},
  url = {https://link.aps.org/doi/10.1103/PhysRev.75.1851}
}

@article{BrownKnepleySmith14,
  author  = {Jed Brown and Matthew G. Knepley and Barry Smith},
  title   = {Run-time extensibility and librarization of simulation software},
  journal = {IEEE Computing in Science and Engineering},
  month   = {Jan},
  volume  = {17},
  number  = {1},
  pages   = {38--45},
  doi     = {10.1109/MCSE.2014.95},
  year    = {2015},
  petsc_uses={KSP},
}

@inproceedings{Byers1970,
   author    = {J.A. Byers},
   booktitle = {Proceedings of the Fourth Conference of Numerical Simulation of Plasmas},
   title     = {Noise Suppression Techniques in Macroparticle Models of Collisionless Plasmas},
   editor    = {NTIS},
   year      = {1970},
   location  = {NRL, Washington, D.C.}
}

@article{Canosa1973,
   title = {Numerical solution of Landau's dispersion equation},
   journal = {Journal of Computational Physics},
   volume = {13},
   number = {1},
   pages = {158-160},
   year = {1973},
   issn = {0021-9991},
   doi = {https://doi.org/10.1016/0021-9991(73)90131-9},
   url = {https://www.sciencedirect.com/science/article/pii/0021999173901319},
   author = {José Canosa}
}

@book{Chen1984,
   title = {Introduction to Plasma Physics and Controlled Fusion},
   author = {F.F. Chen},
   year = {1984},
   publisher = {Plenum Press, New York},
   volume = {1},
   edition = {2}
}

@article{Cheng1976,
   title = {The Integration of the Vlasov Equation in Configuration Space},
   journal = {Journal of Computational Physics},
   volume = {22},
   number = {3},
   pages = {330-351},
   year = {1976},
   issn = {0021-9991},
   doi = {https://doi.org/10.1016/0021-9991(76)90053-X},
   url = {https://www.sciencedirect.com/science/article/pii/002199917690053X},
   author = {C.Z Cheng and Georg Knorr}
}

@Book{Ciarlet1976,
  title={Numerical Analysis of the Finite Element <ethod},
  author={Philippe G. Ciarlet},
  Publisher={Les Presses de L'Universit\'{e} de Montréal},
  year={1976}
}

@article{Dawson1961,
author = {Dawson,John },
title = {On Landau Damping},
journal = {The Physics of Fluids},
volume = {4},
number = {7},
pages = {869-874},
year = {1961},
doi = {10.1063/1.1706419},
URL = {https://aip.scitation.org/doi/abs/10.1063/1.1706419},
eprint = {https://aip.scitation.org/doi/pdf/10.1063/1.1706419}
}

@misc{Denavit1981,
   address = {New York},
   issn = {0374-2806},
   language = {eng},
   lccn = {75640465},
   publisher = {[Gordon and Breach Science Publishers]},
   series = {Comments on Modern Physics, pt. E},
   title = {Comments on Plasma Physics and Controlled Fusion.},
   author = {J. Denavit and J.M. Walsh},
   year = {1988},
}

@inproceedings{Friedberg1969,
  author          = {J.P. Friedberg and R.L. Morse and C.W. Nielson},
  booktitle       = {Proceedings of the Third Conference on Numerical Simulation of Plasmas},
  year            = {1969},
  location        = {Standford University, CA}
}

@Inbook{Hairer2006_2,
author="Hairer, Ernst
and Wanner, Gerhard
and Lubich, Christian",
title="Symplectic Integration of Hamiltonian Systems",
bookTitle="Geometric Numerical Integration: Structure-Preserving Algorithms for Ordinary Differential Equations",
year="2006",
publisher="Springer Berlin Heidelberg",
address="Berlin, Heidelberg",
pages="179--236",
isbn="978-3-540-30666-5",
doi="10.1007/3-540-30666-8_6",
url="https://doi.org/10.1007/3-540-30666-8_6"
}

@article{Hapla2021,
   title={Fully Parallel Mesh I/O Using PETSc DMPlex with an Application to Waveform Modeling},
   volume={43},
   ISSN={1095-7197},
   url={http://dx.doi.org/10.1137/20M1332748},
   DOI={10.1137/20m1332748},
   number={2},
   journal={SIAM Journal on Scientific Computing},
   publisher={Society for Industrial & Applied Mathematics (SIAM)},
   author={Hapla, Vaclav and Knepley, Matthew G. and Afanasiev, Michael and Boehm, Christian and van Driel, Martin and Krischer, Lion and Fichtner, Andreas},
   year={2021},
   pages={C127–C153}
}

@book{Harlow1955,
  title={A Machine Calculation Method for Hydrodynamic Problems},
  author={Harlow, F.H. and Evans, M. and Richtmyer, R.D.},
  series={LAMS (Los Alamos Scientific Laboratory)},
  url={https://books.google.com/books?id=rvM5zQEACAAJ},
  year={1955},
  publisher={Los Alamos Scientific Laboratory of the University of California}
}

@book{Harlow1967,
   title = {The Particle-In-Cell Method for Numerical Solution of Problems in Fluid Dynamics},
   author = {Harlow, Francis H},
   abstractNote = {The particle-in-cell method is a procedure to be used on high-speed computer for studies of the dynamics of compressible fluids undergoing large distortions. The technique is described for calculation of the dynamics of two fluids confined in a two-dimensional rectangular box. Techniques are also discussed for the extension to numerous other types of problems. The properties, limitations, and uses of the method are discussed.},
   doi = {10.2172/4769185},
   url = {https://www.osti.gov/biblio/4769185},
   series={LAMS (Los Alamos Scientific Laboratory)},
   place = {United States},
   year = {1962},
   publisher={Los Alamos Scientific Laboratory of the University of California}
}

@article{Jackson1960,
   doi = {10.1088/0368-3281/1/4/301},
   url = {https://dx.doi.org/10.1088/0368-3281/1/4/301},
   year = {1960},
   publisher = {},
   volume = {1},
   number = {4},
   pages = {171},
   author = {J D Jackson},
   title = {Longitudinal plasma oscillations},
   journal = {Journal of Nuclear Energy. Part C, Plasma Physics, Accelerators, Thermonuclear Research}
}

@article{Kirby04,
  author    = {Robert C. Kirby},
  title     = {Algorithm 839: {FIAT}, a new paradigm for computing finite element basis functions},
  journal   = {ACM Transactions on Mathematical Software},
  volume    = {30},
  number    = {4},
  year      = {2004},
  issn      = {0098-3500},
  pages     = {502--516},
  doi       = {10.1145/1039813.1039820},
  publisher = {ACM Press},
  address   = {New York, NY},
}

@article{KnepleyBrownRuppSmith13,
  author   = {{Knepley}, M.~G. and {Brown}, J. and {Rupp}, K. and {Smith}, B.~F.},
  title    = {Achieving High Performance with Unified Residual Evaluation},
  journal  = {ArXiv e-prints},
  eprint   = {1309.1204},
  archivePrefix = "arXiv",
  primaryClass  = "cs.MS",
  year     = 2013,
  month    = sep,
  adsurl   = {http://adsabs.harvard.edu/abs/2013arXiv1309.1204K},
  petsc_uses={KSP},
}

@article{KnepleyKarpeev09,
  author  = {Matthew G. Knepley and Dmitry A. Karpeev},
  title   = {Mesh Algorithms for {PDE} with {Sieve} {I}: {Mesh} Distribution},
  journal = {Scientific Programming},
  volume  = {17},
  number  = {3},
  pages   = {215--230},
  year    = {2009},
  doi     = {10.3233/SPR-2009-0249},
  url     = {http://arxiv.org/abs/0908.4427},
  note    = {\url{http://arxiv.org/abs/0908.4427}},
  petsc_uses={DMPlex},
}

@eprint{KnepleyLangeGorman2017,
  title     = {Unstructured Overlapping Mesh Distribution in Parallel},
  author    = {Matthew G. Knepley and Michael Lange and Gerard J. Gorman},
  journal   = {arXiv},
  eprint    = {1506.06194},
  year      = {2017},
  petsc_uses={DMPlex},
}

@article{Landau1936,
   author="L.D. Landau",
   title="Die kinetische Gleichung für den Fall Coulombscher Wechselwirkung",
   Journal="Phys. Zs. Sowjetunion",
   volume="10",
   pages="154–164",
   year="1936"
}

@article{Landau1946,
   author="L.D. Landau",
   title="On the Vibrations of the Electronic Plasma",
   Journal="Journal of Physics USSR",
   volume="10",
   pages="25-34",
   year="1946"
}

@article{LangeMitchellKnepleyGorman2015,
  title     = {Efficient mesh management in {Firedrake} using {PETSc-DMPlex}},
  author    = {Michael Lange and Lawrence Mitchell and Matthew G. Knepley and Gerard J. Gorman},
  journal   = {SIAM Journal on Scientific Computing},
  volume    = {38},
  number    = {5},
  pages     = {S143--S155},
  eprint    = {http://arxiv.org/abs/1506.07749},
  doi       = {10.1137/15M1026092},
  year      = {2016},
  petsc_uses={DMPlex},
}

@article{MaddisonFarrell2012,
  title   = {Directional integration on unstructured meshes via supermesh construction},
  author  = {James R Maddison and Patrick E Farrell},
  journal = {Journal of Computational Physics},
  volume  = {231},
  number  = {12},
  pages   = {4422--4432},
  year    = {2012},
}

@inproceedings{MayKnepley2017,
  title     = {{DMSwarm}: Particles in {PETSc}},
  author    = {Dave A May and Matthew G Knepley},
  booktitle = {EGU General Assembly Conference Abstracts},
  volume    = {19},
  pages     = {10133},
  year      = {2017},
  petsc_uses={KSP},
}

@article{McKinstrie1999,
   author = {McKinstrie,C. J.  and Giacone,R. E.  and Startsev,E. A. },
   title = {Accurate formulas for the Landau damping rates of electrostatic waves},
   journal = {Physics of Plasmas},
   volume = {6},
   number = {2},
   pages = {463-466},
   year = {1999},
   doi = {10.1063/1.873212},
   URL = {https://doi.org/10.1063/1.873212},
   eprint = {https://doi.org/10.1063/1.873212}
}

@article{Myers2016,
  doi = {10.48550/ARXIV.1602.00747},
  url = {https://arxiv.org/abs/1602.00747},
  author = {Myers, Andrew and Colella, Phillip and Van Straalen, Brian},
  title = {A 4th-Order Particle-in-Cell Method with Phase-Space Remapping for the Vlasov-Poisson Equation},
  publisher = {arXiv},
  year = {2016},
  copyright = {arXiv.org perpetual, non-exclusive license}
}

@misc{petsc-web-page,
  author = {Satish Balay and Shrirang Abhyankar and Mark~F. Adams and Steven Benson and Jed Brown
    and Peter Brune and Kris Buschelman and Emil~M. Constantinescu and Lisandro Dalcin and Alp Dener
    and Victor Eijkhout and Jacob Faibussowitsch and William~D. Gropp and V\'{a}clav Hapla and Tobin Isaac and Pierre Jolivet
    and Dmitry Karpeev and Dinesh Kaushik and Matthew~G. Knepley and Fande Kong and Scott Kruger
    and Dave~A. May and Lois Curfman McInnes and Richard Tran Mills and Lawrence Mitchell and Todd Munson
    and Jose~E. Roman and Karl Rupp and Patrick Sanan and Jason Sarich and Barry~F. Smith
    and Stefano Zampini and Hong Zhang and Hong Zhang and Junchao Zhang},
  title        = {{PETS}c {W}eb page},
  url          = {https://petsc.org/},
  howpublished = {\url{https://petsc.org/}},
  year         = {2022},
}

@manual{PETScManual,
   author = {Satish Balay et. al.},
   title = {PETSc/TAO Users Manual},
   year = {2022},
   language = {English},
   version = {Revision 3.18},
   organization = {Argonne National Laboratory},
   pagetotal = {310}
}

@article{Pusztay2022,
   author = {Pusztay, Joseph V. and Knepley, Matthew G. and Adams, Mark F.},
   title = {Conservative Projection Between Finite Element and Particle Bases},
   journal = {SIAM Journal on Scientific Computing},
   volume = {44},
   number = {4},
   pages = {C310-C319},
   year = {2022},
   doi = {10.1137/21M1454079},
   URL = {https://doi.org/10.1137/21M1454079},
   eprint = {https://doi.org/10.1137/21M1454079}
}

@article{Shalaby2017,
   doi = {10.3847/1538-4357/aa6d13},
   url = {https://dx.doi.org/10.3847/1538-4357/aa6d13},
   year = {2017},
   publisher = {The American Astronomical Society},
   volume = {841},
   number = {1},
   pages = {52},
   author = {Mohamad Shalaby and Avery E. Broderick and Philip Chang and Christoph Pfrommer and Astrid Lamberts and Ewald Puchwein},
   title = {SHARP: A Spatially Higher-order, Relativistic Particle-in-cell Code},
   journal = {The Astrophysical Journal}
}

@book{SonnendruckerBook,
   author = {E. Sonnendr\"{u}cker},
   title={Numerical Methods for the Vlasov–Maxwell equations (\textit{Book in preperation})}
}

@article{VanKampen1955,
   title = {On the Theory of Stationary Waves in Plasmas},
   journal = {Physica},
   volume = {21},
   number = {6},
   pages = {949-963},
   year = {1955},
   issn = {0031-8914},
   doi = {https://doi.org/10.1016/S0031-8914(55)93068-8},
   url = {https://www.sciencedirect.com/science/article/pii/S0031891455930688},
   author = {N.G. {Van Kampen}}
}

@article{Villani2009,
   author = {Clement Mouhot and Cedric Villani},
   title = {{On Landau Damping}},
   volume = {207},
   journal = {Acta Mathematica},
   number = {1},
   publisher = {Institut Mittag-Leffler},
   pages = {29 - 201},
   year = {2011},
   doi = {10.1007/s11511-011-0068-9},
   URL = {https://doi.org/10.1007/s11511-011-0068-9}
}

@Article{Vlasov1938,
 Author = {A. {Vlasov}},
 Title = {{\"Uber die Schwingungseigenschaften des Elektronengases}},
 FJournal = {{Zhurnal \`Eksperimentalno\u{\i} i Teoretichesko\u{\i} Fiziki}},
 Journal = {{Zh. \`Eksper. Teor. Fiz.}},
 ISSN = {0044-4510},
 Volume = {8},
 Pages = {291--318},
 Year = {1938},
 Publisher = {Academy of Sciences of the Union of Soviet Socialist Republics - USSR (Akademiya Nauk SSSR), Moscow; MAIK ``Nauka/ Interperiodika'', Moscow},
 Language = {Russian},
 Zbl = {0022.18102}
}

@article{Wathen1987,
  title   = {Realistic eigenvalue bounds for the Galerkin mass matrix},
  author  = {Andrew J Wathen},
  journal = {IMA Journal of Numerical Analysis},
  volume  = {7},
  number  = {4},
  pages   = {449--457},
  year    = {1987},
}

@article{ZhouGuoShu2001,
  title = {{Numerical study on Landau damping}},
  author = {Tie Zhou and Yan Guo and Chi Wang Shu},
  journal = {Physica D: Nonlinear Phenomena},
  volume = {157},
  number = {4},
  pages = {322--333},
  doi = {10.1016/S0167-2789(01)00289-5},
  issn = {01672789},
  year = {2001}
}

\end{document}